\documentstyle[aps,amssymb,epsfig]{revtex}
\begin{document}

\title{Computer simulation studies of wetting on heterogeneous surfaces}

\author{
S. Curtarolo$^{1,2}$, M.J. Bojan$^1$, G. Stan$^1$, M.W. Cole$^1$, and W.A. Steele$^{1,3}$} 
\address{
$^1$Department of Physics and Chemistry, Penn State University, University Park, PA 16802, USA\\
$^2$Department of Materials Science and Engineering, MIT, Cambridge, MA 02139, USA\\
$^3${corresponding author e-mail: was@chem.psu.edu}
}
%\date{\today}

\maketitle
\begin{center} 
{\sf Proceeding of 2nd Pacific Basin Conference on Adsorption Science and Technology, \\
May 14-18 2000, Brisbane, Queensland, Australia} 
\end{center} 

\section{Introduction}

The wetting of solid surfaces by fluids is a problem of great practical
importance that has been extensively studied over the years \cite{ref.1}. Most often,
the experimental work has involved measurements of the contact angle $\theta$ made
by a liquid on the solid surface of interest. Macroscopically, this is the
angle made by the surface of a droplet of liquid in contact with a solid
surface and in equilibrium with its vapor. Young's equation gives the
relation between $\cos \theta$ and the interfacial tensions $\gamma_{LV}$, $\gamma_{SL}$ and $\gamma_{SV}$ where
L, S and V denote liquid, solid and vapor respectively:

\begin{equation}
\cos \theta = \frac{\gamma_{SV}-\gamma_{SL}}{\gamma_{LV}}
\label{eq.1}
\end{equation}

(Microscopic expressions relate these surface tensions to spreading
pressure, which is $p_\parallel$, the component of the pressure tensor parallel to the
interface.) The point here is that when the ratio on the right hand side of
eq. (\ref{eq.1}) is larger than one, a droplet will spread into a thin layer (and the
equation actually becomes an inequality, since $\cos \theta$ cannot be greater than
1 or less than -1) and when the ratio is less than -1, the droplet will not
spread. These two regimes are described as wetting and non-wetting
respectively and the regime where $\cos \theta$  is finite is called partial wetting.
Note that the molecular equation that gives interfacial tension as an
integral over the z-dependent local $p_\parallel$ is well established only for those
systems where the fluid-solid interaction is independent of the variables $x,
y$ that give the position of the fluid molecule over the solid. This
limitation is a major obstacle to the use of this formalism in microscopic
calculations of the wetting properties of fluids on rough or heterogeneous
surfaces.

\begin{figure}[b]
% \vspace*{-20mm}
 \centerline{\epsfig{file=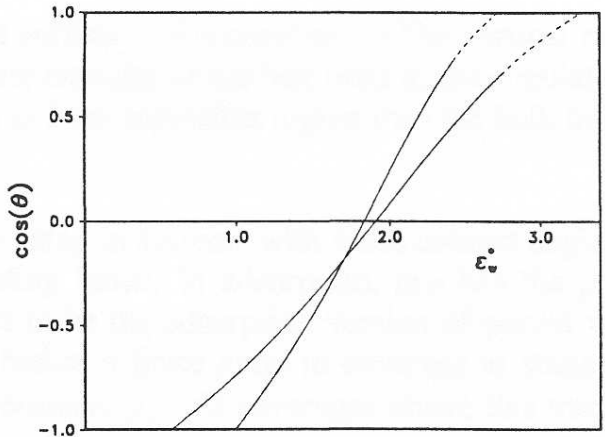,height=6cm}}
 \vspace*{3mm}
 \caption {Cosine of the contact angle for a Lennard-Jones fluid against a
	wall with fluid-wall potential as described in the text. The cosine is shown
	as a function of the ratio $\epsilon_W$ of the well depths for reduced temperatures
	$T^*=kT/\epsilon = 0.5$ and $0.7$ ($0.7$ gives the steeper curve). Here $\epsilon$ is the
	fluid-fluid well depth. The dashed portions of these curves are
	extrapolations of the simulated results. The discontinuous changes in slope
	of $\cos \theta$ at its upper and lower limits mean that the transitions from
	non-wetting or wetting to partial wetting are first order. From Ref. \protect \cite{ref.3}.}
% \vspace*{-1.2cm}
 \label{fig.1}
\end{figure}

However, many molecular simulations that conform to the flat surface
limitation have been reported \cite{ref.2}. An example \cite{ref.3} is shown in Figure 1, where
values of the contact angle are shown for a Lennard-Jones fluid (i.e., one
with potentials that depend upon the inverse 12,6 powers of separation
distance) interacting with a surface with a potential that varies as $z^{-9}$ for
its repulsive part and $z^{-3}$ for its attraction. Results are shown for two
values of temperature $kT/\epsilon$, where $\epsilon$ is the interaction energy well depth for
the fluid-fluid interaction. The quantity $\epsilon_W$ is one of great importance in
the present discussion since it gives the ratio of the fluid-solid well
depth to that for the fluid-fluid interaction. It turns out that the values
of this parameter are generally the most important factor in determining
wetting behavior. Note that the two curves in Figure 1 indicate the ranges
of $\epsilon_W$ that give partial wetting, with complete wetting and drying occurring
for $\epsilon_W$ larger and smaller, respectively, than the interval where $\cos \theta$ is
between 1 and -1.

\begin{figure}[hbt]
 \vspace*{1mm}
 \centerline{\epsfig{file=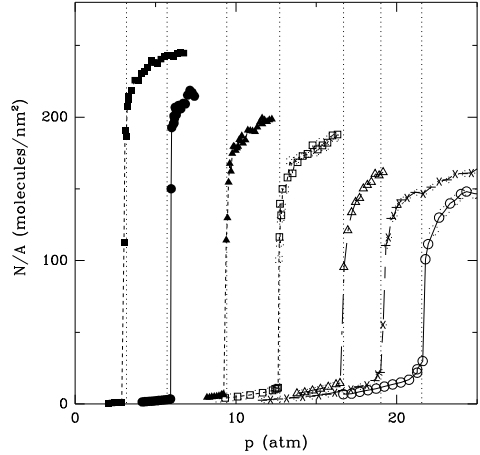,height=6cm}}
 \vspace*{2mm}
 \caption {Isotherms for L. J. neon on a surface with interaction well
   depth=17 K at. $T= 43, 42, 41,39, 37, 34$ and $32$ K., from right to left. All
  are non-wetting. From Ref. \protect \cite{ref.4} }
% \vspace*{-1.2cm}
 \label{fig.2}
\end{figure}

An alternative to the contact angle computation is the more generally
applicable formulation of this problem based on simulations (and
measurements) of adsorption isotherms for a fluid on a weakly interacting
solid surface. In this case, the limits of wetting and non-wetting are
easily stated: wetting is the normal case for a strongly interacting surface
where the adsorbed amount increases smoothly to infinity as the pressure
approaches the bulk vapor pressure of the fluid. The non-wetting case is
also easily described, since the adsorption isotherms in this case never rise
above a rather small level (usually less than a monolayer) until they meet
the bulk vapor pressure line where the container will fill with bulk fluid.
An example \cite{ref.4} is shown in Figure 2, which gives simulated isotherms for
neon adsorbed at several temperatures on a weakly interacting surface. All
of these show a vertical rise at the bulk vapor pressure starting from a
small coverage typical of non-wetting behavior. The estimated monolayer
capacity for neon on this surface is 9 atoms/nm$^2$ . (The vertical rises
terminate at finite values because of the finite capacity of the box used in
the simulations.) The temperature range for these isotherms is from somewhat
higher than the bulk boiling point ($27.1$ K) to close to critical ($44.4$ K).

\begin{figure}[hbt]
 \vspace*{-3mm}
 \centerline{\epsfig{file=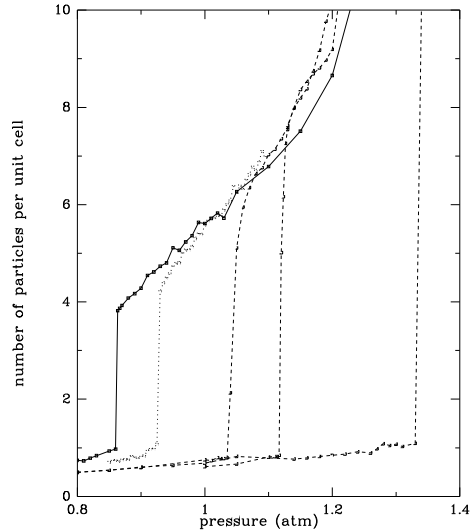,height=7cm}}
 \vspace*{2mm}
 \caption {Isotherms for L.J. neon at $28$ K. on surfaces of varying degree of
	heterogeneity (as described in sec. 3 of the text) The left-hand isotherm
	(solid line) is actually for a homogeneous (flat) surface with well depth
	ratio $\epsilon_W$=0.44. All isotherms show prewetting except for the right hand
	curve, which is non-wetting. Heterogeneities were generated by deleting
	atoms from the surface or, in the non-wetting case, both adding and deleting
	atoms. From. Ref. \protect \cite{ref.5}.}
% \vspace*{-1.2cm}
 \label{fig.3}
\end{figure}

The analogue to partial wetting in systems with finite contact angles is not
as clear as for the wetting and non-wetting limits. In adsorption, one has
the phenomenon known as prewetting which appears to be the adsorption
version of partial wetting. In prewetting, the adsorption isotherm makes a
finite jump in coverage at some value of the pressure less than the bulk
vapor pressure $p_o$. At coverages above this transition, the isotherm is close
to that for a wetting system, with coverages approaching infinity as $p \rightarrow p_o$.
Below the transition, the isotherm has the general appearance of one for a
non-wetting fluid. Examples \cite{ref.5} are shown in Figure 3, which also shows a
single non-wetting isotherm. All are for the temperature of $28$ K
corresponding to a bulk vapor pressure of $1.23$ atm.; the monolayer capacity
is $\approx$ $1.4$ particles per unit cell.

Although the quantitative connection between prewetting isotherms and
partial wetting contact angles has not been established yet, it is evident
that this need not prevent one from using adsorption isotherms to
characterize the wetting behavior of fluids on weakly interacting surfaces.
After giving a brief description of the computer simulation algorithms used
to evaluate the thermodynamic properties of adsorbed films, the remainder of
this paper will be devoted to a presentation of results for simple
Lennard-Jones atoms (rare gases) adsorbed on a number of model weakly
interacting solid surfaces, some of which are heterogeneous.

\section{Computer Simulation Algorithms}

Computer simulation of the macroscopic properties of molecular fluids is now
a very widely used technique to solve the problem of evaluating the
microscopic equations for these properties. These simulations fall into two
general categories: first, molecular dynamics, where the equations of motion
of the molecules are solved numerically to generate their configurations as
a function of time. This data can then be used to evaluate both
thermodynamic and dynamical properties of the system. Second, one can
utilize Monte Carlo algorithms which also generate configurations using a
probabilistic approach. The Monte Carlo algorithms have proved to be very
effective in problems relating to physical adsorption. Therefore the
discussion here will be limited to such computations. We need only briefly
recapitulate the general principles, since these calculations have been
extensively described in monographs \cite{ref.6} and review articles.

In Monte Carlo simulations, atoms are subjected to random moves, which are
accepted or rejected according to rules that will, in the limit of a large
number of trials, produce configurations that correspond to those in the
desired statistical mechanical ensemble. We discuss briefly here the
requirements for generating a canonical ensemble, and then the more useful
Grand Canonical ensemble. A canonical ensemble consists of a collection of $N$
particles in a box of volume $V$ at temperature $T$. Interaction potentials
between particles are chosen so that one can evaluate the total potential
energy of the entire system of particles. Usually, pair-wise interactions
are specified and for atoms, the energy depends only upon their position
coordinates in the box. The system is extended by using periodic boundary
conditions and the minimum image convention, which means that the computer
box of several hundred or thousand particles is surrounded by images of
itself and the box walls are made transparent so that atoms can pass out
(and images pass in). Each atom interacts with its nearby real neighbors as
well as with nearby images. The crucial question is: when does one accept (or
reject) a move than produces a change $\delta U$ in the potential energy of the
system? The answer is in two parts: If $\delta U$ is negative, the system is more
stable and the move should be accepted; if it is positive, then it may or
may not be accepted depending upon the value of $\exp (-\delta U/kT)$. If this is
larger than a randomly chosen number between 1 and 0, the move will be
accepted. Otherwise, it is rejected and another random move is tried. After
a large number of such trials, the configuration of the atoms in the system
will conform to that of the atoms in a canonical ensemble and one can use
such configurations to evaluate averages of quantities such as $<U>$, the mean
potential energy. Note that the mean kinetic energy is $3NkT/2$ in such a
system. The average pressure can also be evaluated using the virial theorem;
more to the point, the elements of the pressure tensor parallel and
perpendicular to the surface in an inhomogeneous fluid such as an adsorbed
layer can also be evaluated.

This brief discussion of a canonical ensemble Monte Carlo has been given
primarily to help establish the ground rules for the Grand Canonical 
ensemble Monte Carlo (GCMC) of a physically adsorbed gas in which the 
quantities to be evaluated are $<U>$ and $<N>$, the average potential 
energy of the adsorbed fluid and the average
number of particles in a system with fixed volume $V$, temperature $T$ and
chemical potential $\mu$. For an adsorbed layer in equilibrium with gas at
pressure $p$, the chemical potentials of the two phases must be equal and,
assuming that the gas phase is ideal (or that corrections to $\mu_{gas}$ for
non-ideality of the gas phase can be made), one has

\begin{equation}
\mu_{gas} = \mu_{ads} = kT \log p + q
\label{eq.2}
\end{equation}

where q is a constant calculable from the standard expression for an atomic
gas. Thus, the GCMC simulation treats $<N>$ and $<U>$ as quantities to be
evaluated from the simulation at fixed $V$, $T$, $\mu_{ads}$ , with $\mu_{ads}$ related to
the observable pressure by eq. 2. The moves in this Monte Carlo procedure
are of three kinds: translation, as in the canonical case; creation of a new
atom at a randomly chosen position in the computer box; and deletion of one
of the pre-existing atoms. The decision concerning the acceptability or not
of a given move now depends upon the value of $\delta C$ or $\delta D$ rather than $\delta U$. $C$ and
$D$ refer to creation or destruction, respectively and are equal to:

\begin{equation}
\delta C=\delta U+\log (zV/(N+1))
\label{eq.3}
\end{equation}

\begin{equation}
\delta D=\delta U + \log (N/zV)
\label{eq.4}
\end{equation}

where N is the instantaneous number of particles in the system and z is the
activity defined by $\exp(\mu/kT)/\Lambda^3$ , where $\Lambda=h /\sqrt{2 \pi m kT}$. (For an ideal gas,
$z=p/kT$.) By combining moves of these three types, one can generate
configurations in the Grand ensemble after a considerably larger number of
moves than for the analogous canonical ensemble. At this point, the isotherm
is essentially a plot of $<N>$ versus $kT \log p= \mu_{ads} +$ constant.

\begin{table}[h]

\caption{Well depth ratios for rare gases on various weakly interacting
solids (From. ref. 4). The gas-gas well depths are listed in parentheses; 
and the theoretical values of $\epsilon_W$, the well depth ratio, are 
listed for the gas-solid pairs indicated.}
\vspace*{0.2 in}
\begin{center}{\begin{tabular}{|c|c|c|} 
Substrate	&Neon		& Argon \\
		&(33.9 K) 	&(120 K)  \\\hline        
Mg      	&2.8 		&3.5  \\\hline 
Li      	&1.5 		&2.0  \\\hline 
Rb      	&0.7 		&1.1  \\\hline 
Cs      	&0.7 		&1.1  \\\hline 
CO$_2$	        &		&2.6  
\end{tabular}}
\end{center}
\label{table.1}
\end{table}
          
\begin{figure}[b]
 \vspace*{-5mm}
 \centerline{\epsfig{file=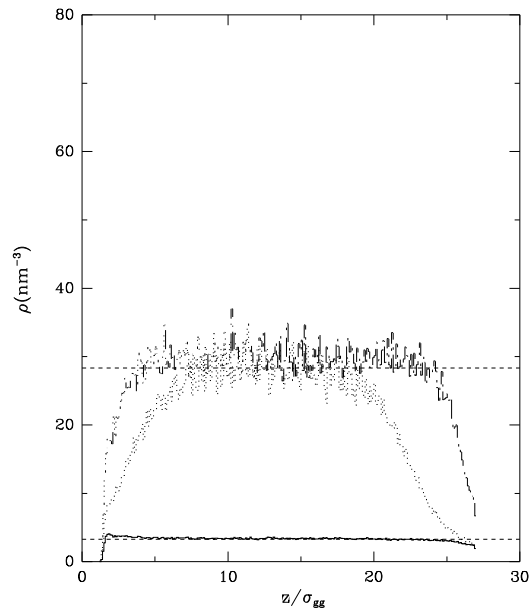,height=8cm}}
 \vspace*{2mm}
 \caption {Local densities in atoms per nm$^3$ for neon at $39$ K in contact with
	a surface with well depth $=17$ K. The densities are plotted versus distance
	$z/\sigma_{gg}$ where $\sigma_{gg} =0.278$ nm. The three curves are for amounts per nm$^2$ of 11
	(solid line at the bottom), $137$ (dotted line) and $183$ (dot-dashed line -
	highest curve). The decreases on the right hand side are due to interactions
	with the hard wall boundary at $z/\sigma_{gg}=27$. The decreases in density when the
	neon is in contact with the adsorbing wall on the left show that the gas
	does not adsorb on the wall for this non-wetting system. The liquid density
	shown by the dashed line indicates that the fluid in the center of the box is
	bulk. From Ref. \protect \cite{ref.4}. }
% \vspace*{-1.2cm}
 \label{fig.4}
\end{figure}

This formalism is very widely used in simulations of adsorption, including
those for weakly interacting solid where wetting is an important issue. The
wetting of polar molecules on non-polar adsorbing surfaces has been studied
for many years. The simpler case of rare gases on real weakly adsorbing
surfaces did not receive much attention until it was pointed out that the
alkali metals should have gas-solid interactions that are sufficiently weak
to give interesting wetting behavior. The well depth ratios for several of
these systems are listed in Table I. Both experiment and theory indicate
that many interesting features are exhibited by these systems. 

\clearpage

The remainder of this paper will be devoted to a description of some of the
results obtained for the wetting behavior of neon adsorbed on these weakly
interacting solid surfaces. Several simulations have been reported for a 
solid called CO$_2$ \cite{ref.4}- a ratio is listed for this material as
well, although one should not take its description as solid CO$_2$ seriously.
From Table I, it is evident that the alkali metals Cs and Rb are expected to
be the weakest possible adsorbents for rare gases. A number of wetting
studies have been carried out for Neon on these materials \cite{ref.4,ref.5}, primarily
because one hopes that neon is sufficiently heavy to obey classical
statistical mechanics, which is the basis for the GCMC algorithms used.

\section{Simulation Results}

Simulations were first performed for the energetically flat approximations
to the gas-solid interactions because of their simplicity. Figures 2 and 3
already show some simulations of neon on model alkali metal surfaces. Note
that the simulations were not restricted to interaction well depths
predicted by theory. Trends in the adsorption behavior could most easily be
observed by taking various interaction strengths that do not necessarily
correspond to the theoretical predictions for a given metal.

\begin{figure}[hbt]
% \vspace*{-20mm}
 \centerline{\epsfig{file=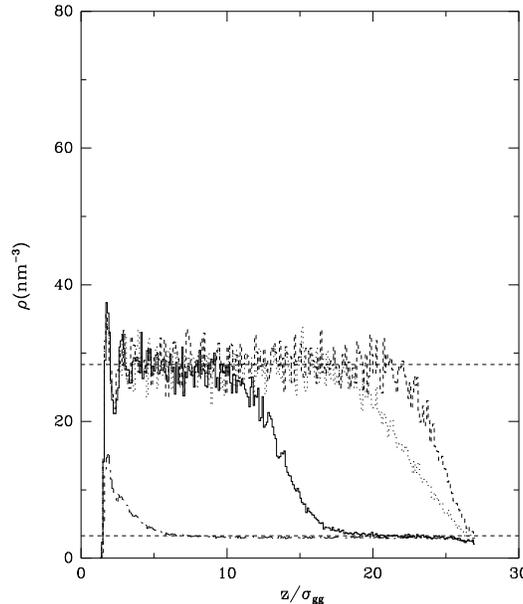,height=8cm}}
% \vspace*{3mm}
 \caption {Same as Fig. 4, but for a gas-solid interaction well depth of $60$
K. This wetting system shows the sharp monolayer peaks at small $z/\sigma_{gg}$
expected for adsorption. The amounts adsorbed (in atoms per nm$^2$) for
increasing densities are $15, 93, 152$ and $170$. From Ref. \protect \cite{ref.4}.}
% \vspace*{-1.2cm}
 \label{fig.5}
\end{figure}

Simulations for Ne on a surface with $\epsilon_W$ = 0.5 are shown in Fig. 2 and
indicate that this system is similar to the Cs case of Table I in that they
also are non-wetting over a range of temperature up to a couple of degrees
below critical. (Simulations closer to critical than these were not feasible
for the usual reason: correlation lengths were beginning to be larger than
the computer box and were therefore introducing artifacts in the isotherms.)
We note that one can readily extract local densities in the adsorption box
from the configurations of the atoms in it. In this work, a box of lateral
dimensions = $2.78 \cdot 2.78$ nm$^2$ and height 7.5 nm was used. In the lateral
directions, periodic boundary conditions were applied; in the "vertical" or
$Z$ direction, the adsorbing surface was a wall at $Z$=0 and a hard wall was
imposed on the opposite side of this box. The local densities are shown in
Figure 4 for a non-wetting case, in Figure 5 for a wetting example and in
Figure 6 for a prewetting system. For a non-wetting system, the figure shows
that the densities of atoms at both the lower and upper walls is very small,
as one might expect. (This behavior has been observed previously by Adams
and Henderson \cite{ref.3}). Thus, no complete monolayer is formed, giving very small
adsorption over the entire pressure range up to $p_o$. If the gas-solid well
depth is increased, one observes prewetting, as in Figure 6. The densities
for this prewetting case are at least qualitatively similar to those for the
wetting system shown in Fig. 5. Note that this prewetting transition
involves a jump in coverage from about $2.2$ atoms per nm$^2$ (0.24 layer) to
30.3 atoms per nm$^2$ ($3.3$. layers) for $\epsilon_W$=2.8 at T=28 K.The prewetting regime
for a solid with well depth of $95$ K ranges from $\approx$ $22$ K to $\approx$ $29$ K.These
results appear to depend somewhat upon the width of the potential well in
addition to its depth \cite{ref.4}.

As for the explicitly heterogeneous surface, we start from the assumption
that any variation in gas-solid energy across the surface can be described as
heterogeneity. Thus the periodic energy variations due to the effect of
atomic structure when the pair-wise additive approximation for the
adsorption potential replaces the integrated perfectly flat surface falls
into this category and thus were briefly considered \cite{ref.5}. Figure 7 shows that
the isotherm obtained for a periodic surface with average interaction energy
equal to that

\begin{figure}[hbt]
% \vspace*{-20mm}
 \centerline{\epsfig{file=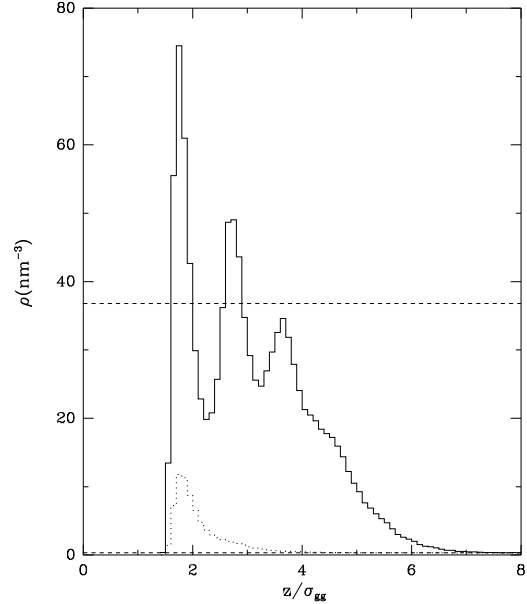,height=8cm}}
 \vspace*{3mm}
 \caption {Densities on a much-expanded distance scale compared to Figs. 4
and 5. Neon at 28 K is adsorbed on a surface with gas-solid well depth = 95
K. The system exhibits prewetting and the densities shown here correspond to
coverages just before the prewetting jump (2.2 atoms/nm$^2$) - the lower curve
- and just after (30.3 atoms/nm$^2$ ).The bulk liquid density is shown by the
dashed horizontal line. The structure of the peaks obtained in this
simulation is quite similar to that expected for a normal wetting system..
From. Ref. \protect \cite{ref.4}.}
% \vspace*{-1.2cm}
 \label{fig.6}
\end{figure}

\begin{figure}[hbt]
% \vspace*{-20mm}
 \centerline{\epsfig{file=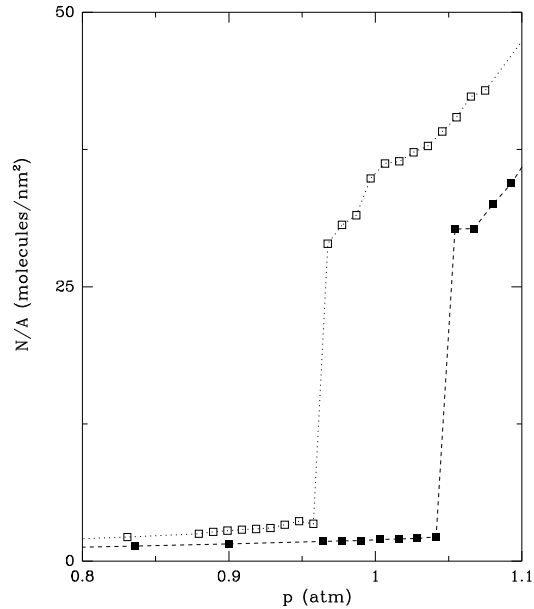,height=8cm}}
 \vspace*{3mm}
 \caption {Simulated adsorption isotherms for neon at $28$ K on a surface with
well depth = $95$ K. The open squares show the isotherm for the surface with a periodic part
(wave length = 0.40 nm) oscillating about the value for the flat surface
(filled squares). Both surfaces appear to show first order wetting
transitions. At this temperature, $p_o =1.22$ atm. From. Ref. \protect \cite{ref.4}.}
% \vspace*{-1.2cm}
 \label{fig.7}
\end{figure}

for the flat surface is noticeably different from that for the flat surface.
What has happened is that the Boltzmann-weighted average gas-solid
interaction for the periodic case is somewhat more negative than that for
the flat surface, giving a Henry's Law low coverage isotherm that is slightly 
steeper for the periodic case than for
the flat. In particular, the prewetting jumps in Fig. 7 start from coverages
of $0.35$ monolayers (periodic) and $0.24$ monolayers (flat) and end essentially
at $3.0$ monolayers.

\section{Conclusions}

The two approaches to wetting based on Young's equation for the surface
tensions and adsorption isotherms should be mutually consistent. It is
likely that the boundaries for the surface tension wetting$\rightarrow$partial wetting
and partial wetting$\rightarrow$non-wetting are indeed consistent with the isotherm
results, if one assumed that prewetting is in fact an alternative notation
for partial wetting. (Note also that non-wetting here means only slight
adsorption before the jump to bulk phase condensation occurs, which has
elsewhere been described as partial wetting, thus confusing the issue.) The
points of resemblance between partial wetting from Young's equation and
prewetting from the isotherms have not been elucidated. Indeed, no detailed
comparison of the wetting and non-wetting boundaries as a function of $T^*$ and
$\epsilon_W$ for the two methods has yet been presented.

Not many simulations of isotherms showing prewetting have been presented
until fairly recently. This is at least in part because the prewetting jumps
appear to occur at pressure very close to the bulk condensation value. This
not need be the case, since it has been shown that prewetting isotherms on
surfaces with some roughness, either due to periodic variation in energy
across the surface (see Figure 7) or to heterogeneity introducing by adding
or deleting atoms from the surface to form defects will cause the prewetting
jump to shift down to a pressure where the transition is more easily
observed. However, for sufficient energy defects in the surface, the
transition is no longer sharp; i.e., not first order. Still, the observation
of truly first order transitions in computer simulations has always been
difficult The situation here where one has no hint of where the first order
behavior does occur other than from the simulated isotherms (or from the
shape of $\cos \theta$ curves such as those shown in Fig. 1) is evidently one where
the difficulty of determining first order transitions is a major one.

\end{document}